\documentclass[%
reprint,superscriptaddress,
amsmath,amssymb,aps,floatfix,prb
]{revtex4-2}
\usepackage{lipsum}
\usepackage[hidelinks]{hyperref}
\usepackage{orcidlink}
\usepackage{graphicx}
\usepackage{dcolumn}
\usepackage{bm}% bold math

\usepackage{tikz}
\usepackage{pgfplots}
\usepackage{pgfplotstable}
\usepgfplotslibrary{groupplots}
\pgfplotsset{compat=1.11}
\usepackage{xcolor}
\usepackage[mathlines]{lineno}% Enable numbering of text and display math
\newcommand{\D}{\mathcal{D}}
\newcommand{\Tr}[0]{\text{Tr}}

\newcommand{\ket}[1]{\left| #1 \right \rangle}
\newcommand{\bra}[1]{\left \langle #1 \right|}
\newcommand{\braket}[2]{\left \langle #1 \middle| #2 \right \rangle}
\newcommand{\braketmatrix}[3]{\left \langle #1 \middle| #2 \middle| #3 \right \rangle}

\begin{document}

%\preprint{APS/123-QED}

\title{Quantum vs thermal fluctuations in phase transitions of two-dimensional superconductors}

\author{A. Ponticelli \orcidlink{0009-0004-8987-5854}}
 \email{andrea.ponticelli@unina.it}
\author{F. G. Capone \orcidlink{0009-0000-9832-4385}}
\affiliation{%
 Dip. di Fisica E. Pancini - Università di Napoli Federico II - I-80126 Napoli, Italy
 }
\affiliation{%
INFN, Sezione di Napoli - Complesso Universitario di Monte S. Angelo - I-80126 Napoli, Italy
}%
\author{V. Cataudella \orcidlink{0000-0002-1835-1429}}
\author{G. De Filippis \orcidlink{0000-0003-0557-3556}}
\author{A. de Candia \orcidlink{0000-0002-9869-1297}}
\author{C. A. Perroni \orcidlink{0000-0002-3316-6782}}
\affiliation{%
INFN, Sezione di Napoli - Complesso Universitario di Monte S. Angelo - I-80126 Napoli, Italy
}%
\affiliation{%
 SPIN-CNR and Dip. di Fisica E. Pancini - Università di Napoli Federico II - I-80126 Napoli, Italy }

%\date{\today}

\begin{abstract}
We investigate the impact of quantum and thermal phase fluctuations on the suppression of superconducting order in two-dimensional systems. Within the two-dimensional quantum XY model  in the phase representation, where on-site interaction terms govern quantum phase fluctuations, we perform extensive path-integral quantum Monte Carlo simulations. The resulting temperature–interaction phase diagram establishes the presence of a well-defined critical line ending at a quantum critical point at vanishing temperature with no indication of reentrant behavior. We further demonstrate that the resistance above the critical line reproduces the two expected different critical behaviors. For stronger interactions, above the quantum critical point, the system exhibits a crossover to an insulating regime at low temperatures. Finally, Monte Carlo calculations of current–current correlation functions enable us to extract the frequency-dependent conductivity in both superconducting and normal regimes, revealing a finite-frequency response that we attribute to quantum phase fluctuations.
%We investigate the role of   both quantum and thermal fluctuations  %in reducing the superconducting phase of two-dimensional systems. 
%The two-dimensional quantum XY model in the phase representation, %where on-site interaction terms control quantum phase fluctuations, 
%is extensively studied via path integral quantum Monte Carlo %techniques. A phase diagram of temperature vs. interaction energy is %thoroughly carried out confirming that there is a well defined %quantum critical point and  no evidence of a reentrant phase %transition. We also show that a well known  critical thermal %behaviour for the resistance is fully reproduced up to the critical 
%value of the interaction energy, while, for larger values of it, we %find the onset of an insulating phase at low temperatures. Finally, %Monte Carlo estimation of the current-current correlations allows to %evaluate the frequency dependent conductivity  in both %superconducting and normal regions, getting evidence of a response %at finite frequency that is ascribed to quantum fluctuations.
\end{abstract}

\maketitle

%\tableofcontents

\section{\label{sec:intro}Introduction}
The classical XY model and its quantum (QXY) extension  describe phase fluctuations \cite{Sachdev,Phillips} in  two dimensional (2D) systems such as superconducting films~\cite{Halperin1979}, arrays of Josephson junctions~\cite{jja-bkt} and superfluid surfaces~\cite{superfluid-surfaces}. More recently, the observation of a superconducting phase at the interface between different oxides, such as LaAlO$_3$/SrTiO$_3$~\cite{Caviglia2008} or LaAlO$_3$/KTaO$_3$~\cite{Mallik2022}, has renewed interest in the 2D QXY model.

The classical 2D XY model describes planar spins that can point in any direction within the plane. In contrast with the three dimensional (3D) case, thermal fluctuations in two dimensions are strong enough to prevent the formation of true long-range order at any finite temperature \cite{Chaikin,Kardar}. Indeed, the 2D model exhibits the Berezinskii–Kosterlitz–Thouless (BKT) transition~\cite{40BKT}: at low temperatures the system enters a quasi-ordered phase characterized by algebraically decaying correlations, while at higher temperatures unbinding of vortex–antivortex pairs destroys this quasi-order. The essential physics of the classical model in two dimensions is therefore governed by the interplay between spin-wave fluctuations and vortices.

%The classical model undergoes a continuous phase transition, the so called %Berezinskii–Kosterlitz–Thouless  transition~\cite{40BKT}, that well describes the %transition observed in those systems.  

The 2D QXY model incorporates,  in addition to thermal fluctuations, quantum ones which can be present even at zero temperature and can significantly modify the phase diagram and the nature of correlations. Indeed, the system  undergoes a quantum phase transition at zero temperature~\cite{Sachdev} that has no analogue in the classical version, reflecting the competition between coherent couplings and quantum fluctuations. At finite temperature, the combined effect of thermal and quantum fluctuations shapes the crossover between quantum-dominated behavior at low temperatures and classical BKT-like physics at higher temperatures. 

Although XY and QXY models have been studied with many different approaches (see, for instance, review articles~\cite{Minnhagen_rev, Fazio_rev}), in our opinion, a comprehensive study based on an unbiased and numerically exact approach—one that provides a unified perspective on both thermodynamics and conductivity~\cite{Halperin1979,Halperin2}—is still lacking. In particular, it is important to clarify the role of quantum fluctuations on the frequency-dependent conductivity  not only close to the quantum critical point~\cite{Sachdev,Sachdev2}, but also far from it analyzing all the range of system parameters.  

%in our opinion it is still lacking a comprehensive study based on an unbiased and %numerically exact approach giving a unified view of both thermodynamics and, in %particular, frequency dependent conductivity. 

\begin{figure}[!h]
    \centering
    \includegraphics[width=\linewidth]{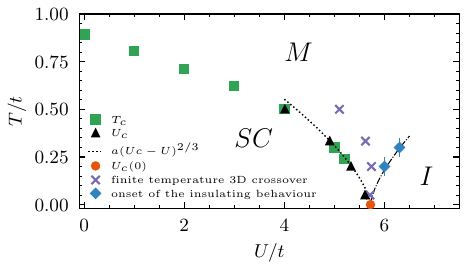}
    \caption{Phase diagram in the $T-U$ plane. \textit{SC} denotes the superconducting, \textit{M} the metallic , and \textit{I} the insulating phase. The temperature $T$ and the interaction energy $U$ quantifies the strength of thermal and quantum fluctuations, respectively, while $t$ is the Josephson coupling energy, used as reference (see \hyperref[sec:model]{Sec.~\ref{sec:model}}). The diagram reports the phase-transition points extracted via finite-size scaling (\hyperref[sec:finite-size-scaling]{Sec.~\ref{sec:finite-size-scaling}}) of the Monte Carlo simulation data (\hyperref[sec:methods]{Sec.~\ref{sec:methods}}). Green squares mark the BKT critical temperatures $T_c=T_c^{BKT}$ obtained at fixed $U$, whereas black triangles indicate the BKT critical couplings $U_c^{BKT}$ determined at fixed temperature (\hyperref[sec:finite-size-scaling]{Sec.~\ref{sec:finite-size-scaling}}). The orange circle marks the quantum critical value $U_c=U_c^{3D}(T=0)$, which belongs to the 3D-XY universality class. The dotted curve emanating from the quantum critical is a fit of the points near the $T=0$ quantum phase transition, obtained using the 3D-XY power-law indicated in the figure (\hyperref[sec:3d-xy]{Sec.~\ref{sec:3d-xy}}). The purple crosses indicate the pseudo 3D-XY phase transition (\hyperref[sec:fake_3d]{Sec.~\ref{sec:fake_3d}}). The blue diamonds denote the  resistance minimum that signals the M–I crossover from metallic to insulating behaviour (the dashed line is only a guide for the eye) (\hyperref[sec:corr-len]{Sec.~\ref{sec:corr-len}}).}
    \label{fig:diagram}
\end{figure}

In this paper, we employ the Path Integral Quantum Monte Carlo (PIQMC) method~\cite{Suzuki1993, suzuki} to perform a detailed study of the QXY model. We use its phase representation, in which on site interaction term, $U$, controls the strength of quantum phase fluctuations.
In Fig.~\ref{fig:diagram} we anticipate the phase diagram that will be described and discussed in the following sections. The results agree with the consolidated understanding of the 2D QXY model, confirming the existence of a clearly defined quantum critical point at zero temperature ~\cite{Sachdev,jose_phasediag} and the absence of a re-entrant phase transition. In any case, to the best of our knowledge, 
it is the first time that all the phases have been identified with high precision within a single study. Actually, in Fig.~\ref{fig:diagram}, we have evidence of the superconducting ($SC$), metallic ($M$) and insulating ($I$) phases. Moreover, the critical line between the metallic and superconducting phases has the $BKT$ critical behaviour at any temperature except at $T=0$, where the critical point takes place at $U_c=U_c(0)$ with the critical exponents of the 3D  XY model (see Fig.~\ref{fig:diagram}). We also have evidence of a pseudo-phase transition just above the metallic-superconducting critical line (violet cross points in Fig.~\ref{fig:diagram}), where the 3D XY fluctuations survive up to finite size.

We also demonstrate that, up to the critical value $U_c$, the temperature dependence of the resistance is fully described by Halperin and Nelson formula~\cite{Halperin1979}, while, for bigger values of interaction, we observe the beginning of an insulating phase at low temperatures.
Lastly, the frequency-dependent conductivity in both superconducting and normal regions can be evaluated using Monte Carlo estimation of the current-current correlations, providing evidence of a response at finite frequency that is attributed to quantum fluctuations.

The paper is organized in the following way: in section II the $QXY$ model is introduced and the Monte Carlo method is briefly discussed, in section III the thermodynamic properties are presented providing the evidences supporting the construction of the phase diagram, finally section IV is devoted to the study of conductivity properties. We report details on the partition function in appendix A and on the linear response theory in appendix B.

\section{\label{sec:model}Model and methods}
The 2D QXY model in the phase representation  can be derived from the repulsive Bose-Hubbard model \cite{BHmodel} in the following form:
\begin{equation}\label{eq:bosehubbard}
    H =  -t \sum_{\langle ij\rangle} \cos (\phi_i-\phi_j) +  \frac{U}{2}\sum_i \left(-i \frac{\partial}{\partial \phi_i}  \right)^2 
\end{equation}
where $i,j$ indicate the sites of an $L\times L$ square lattice, whose nearest-neighbour bonds are denoted by  the symbol ${\langle ij\rangle}$, and $\phi_i$ is the phase at the site $i$. The first term of the Hamiltonian is proportional to the Josephson coupling energy, $t$, between nearest-neighbour sites, which favours phase coherence and it is proportional to the density of the Cooper pairs. The second term of the Hamiltonian describes the quantum effects, quantified by the interaction energy $U$, which lead to phase fluctuations even at zero temperature. Since the model  focuses on the role of phase fluctuations, it is valid well below the mean field critical temperature.

%\begin{equation}\label{eq:bosehubbard}
%    H_{BH} = \frac{U}{2}\sum_i \hat{n}^2_i + t_0 \sum_{\langle ij\rangle}%(\hat{b}^\dagger_i \hat{b}_j + H.c)
%\end{equation}
%where $i,j$ are the spatial site index, $\hat{b}_i$, $\hat{b}^\dagger_i$ are the %annihilation and creation bosonic operators: $[\hat{b}_i, \hat{b}^\dagger_j]=\delta_{i,j}$ %and $\hat{n}_i=\hat{b}^\dagger_i \hat{b}_i$ are the number cooper pairs operators on site. %The first term of the Hamiltonian represents the on-site repulsion due to the charge %density, with $U$ being the strength of the Coulomb interaction. The second term %corresponds to the hopping energy, $t_0$, between nearest-neighbour sites. We use the %phase-amplitude representation \cite{phase-amplitude-BH}, %$\hat{b}_j=\sqrt{n_0}e^{i\hat{\phi}_j}$, where $n_0$ is the most probable number %occupation value for single site, i.e., this approximation neglects the fluctuations of %the site occupation. 

Using the Suzuki-Trotter approach \cite{suzuki} to compute the partition function and the Villain approximation \cite{Villain1975}, the model can be mapped onto an effective classical model (see Appendix \ref{apx:part.func}) with the following Euclidean action:
\begin{equation}
    \begin{gathered}
        S[\{\phi(\tau)\}]=-\sum_{\tau, \langle ij \rangle} K_{s} \cos{\left(\phi_i(\tau)-\phi_j(\tau)\right)} \\
        -\sum_{\tau, i} K_{\tau} \cos{\left(\phi_i(\tau)-\phi_i(\tau+\Delta\tau)\right)},
    \end{gathered}
    \label{eq:action}
\end{equation}
where $\Delta\tau$ denotes the imaginary-time discretization, with $N_\tau = \beta/\Delta\tau$ and $\tau \in \{0,\Delta\tau,2\Delta\tau,\ldots,(N_\tau-1)\Delta\tau\}$. Moreover, $K_s = t\Delta\tau$ and $K_\tau = 1/(U\Delta\tau)$. Consequently, the action is a functional of the paths $\{\phi(\tau)\} = (\{\phi(0)\},\{\phi(\Delta\tau)\},\ldots,\{\phi((N_\tau-1)\Delta\tau)\})$, subject to the periodic boundary condition $\{\phi(0)\} = \{\phi(\beta)\}$. According to this notation, $\{\phi(k\Delta\tau)\}$ represents a phase configuration of the 2D system at imaginary time  $\tau = k\Delta\tau$. In this framework, the original 2D model has been mapped in an effective three dimensional action, where the third dimension, given by $\tau$, naturally introduces the quantum phase fluctuations quantified by the second term in Eq.~\eqref{eq:action}. 
Since the Hamiltonian depends on the two parameters, $t$ and $U$, we choose $t$ as the energy unit. Moreover, throughout this work, we set $\hbar = k_B = 2e = 1$, with $\hbar$ Planck constant, $k_B$ Boltzmann constant, and $e$ electron charge.

In order to analyze the model properties, it is useful to introduce the current  operator $I$ in the phase representation
\begin{equation}\label{eqn:current_operator}
    I = \left[\sum_i \frac{\partial}{\partial\phi_i},\, H \right]
      =  t \sum_{\langle ij\rangle} \sin(\phi_i - \phi_j),
\end{equation}
which, in the path-integral formulation, takes the form
\begin{equation}
    I =  t\, \frac{\Delta\tau}{\beta} \sum_{\langle ij\rangle,\, \tau}
        \sin\!\left[\phi_i(\tau) - \phi_j(\tau)\right].
        \label{eq:path-cur}
\end{equation}

%, so $K_{s} = \Delta\tau$, and $K_{\tau} = t/U\Delta\tau$, and fixed $t=1$.

\subsection{\label{sec:methods}Methods}
Average values of the observables, including the correlation functions calculated in this paper, have been obtained  performing  PIQMC simulations~\cite{barker}  with the Wolff~\cite{wolff, Jiang2022} and Metropolis~\cite{Metropolis} update schemes. The Wolff algorithm exploits the $O(2)$ symmetry of the $XY$ model: the global move is built on the reflection of the cluster grown around a fixed axis for each move with  the following  probability
\begin{equation}
\begin{gathered}
    P_{\text{add}}(\phi_1, \phi_2)=\\
    1-\exp{\left\{ \min{\left[ 0,\ -2K\cos{(r-\phi_1)}\cos{(r-\phi_2)} \right]} \right\}},
\end{gathered}
\end{equation}
where $r$ is the angle with respect to the line orthogonal to the reflection axis, $K$ is $K_\tau$, if the direction is along the imaginary time, or $K_s$ if the direction is along the 2D spatial lattice. The addition of the Metropolis move serves to avoid the slowing down near the critical point, where the Wolff algorithm, due to the long-range correlations, creates a cluster with a size comparable to the system.

\section{\label{sec:results}Thermodynamic properties}
In this section we discuss several thermodynamic properties obtained starting from the partition function (for details see Appendix A).  
In subsection \ref{sec:stiffness}, we analyze the superfluid stiffness at finite size. In subsection \ref{sec:finite-size-scaling}, we exploit the finite size scaling to extract thermodynamic values via Monte Carlo simulations.  In subsection \ref{sec:vortex}, we discuss the role played by vortices calculating their density.
Finally, in subsection \ref{sec:corr-len}, we analyze the correlation length and its relation to the resistance.

\subsection{\label{sec:stiffness}Superfluid stiffness}
\begin{figure}[!ht]
    \centering
    \includegraphics[width=\linewidth]{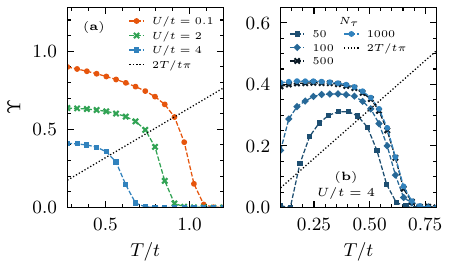}
    \caption{Stiffness $\Upsilon$ as a function of the temperature $T$ for the size $L=48$. (a) Results  are shown for the interaction strengths $U/t=0.1,2,4$ at the discretization number $N_\tau$=1000. The dotted line indicates the NK curve following the criterion given in Eq. (\ref{eq:stiff-scal}), which is typically used to mark the finite temperature transition points.  (b) Results are displayed for discretization numbers $N_\tau=50,100,500,1000$ at the interaction energy $U/t=4$. }
    \label{fig:stiff}
\end{figure}

In order to build up the $T-U$ phase diagram  shown in Fig. \ref{fig:diagram}, we have calculated the temperature dependence of one of the most relevant thermodynamic properties, the superfluid stiffness $\Upsilon$~\cite{superfluid.fisher, superfluid.otha} 
%\begin{equation}
%    \rho_s(T, U) = (m/\hbar)^2\Upsilon(T, U)
%\end{equation}
%where

\begin{equation}
    \Upsilon(T, U) = \frac{1}{2t}\frac{\partial^2 f(T, U, \Phi_0)}{\partial \Phi_0^2}\bigg|_{\Phi_0=0} = \frac{1}{t}
    \frac{\langle S_s \rangle - \beta\langle I^2 \rangle}{2 L^2} \label{eq:stiffness}
\end{equation}
where $f$ is the free energy, and $\Phi_0$ is the boundary phase twist. In Eq.~\eqref{eq:stiffness}, within the path-integral formulation,  $S_s=\frac{1}{\beta}\sum_{\langle ij \rangle,\tau} K_{s} \cos(\phi_i(\tau)-\phi_j(\tau))$, and $I$ is the current defined in Eq.~\eqref{eq:path-cur}. A non zero value $\Upsilon$ is usually used to signal the transition to a superconducting phase. Actually, it has been proved that, in the thermodynamic limit ($L\mapsto \infty$), $\Upsilon$ jumps to zero at the critical temperature with a universal critical jump which can be considered the fingerprint of the BKT transition. Fig.~\hyperref[fig:stiff]{\ref{fig:stiff}(a)} shows the curves $\Upsilon$ as a function of the temperature  at fixed $U$ for a finite size $L$. As expected, finite size effects smooth the critical jump, but there is still a clear evidence that $\Upsilon$ goes to zero with increasing temperature signalling the existence of different phases. In section~\ref{sec:finite-size-scaling} we will discuss how the critical temperature can be extracted in an effective way overcoming the size effects. In Fig.~\hyperref[fig:stiff]{\ref{fig:stiff}(a)} we also observe that, with increasing $U$, the superconducting phase is obtained at lower temperatures, i.e. quantum fluctuations reduce the superconducting stability. 

%Increasing quantum fluctuations leads to a suppression of the stiffness —and thus the %superconductivity density— resulting in a reduced critical temperature.

Since the mapped model is obtained by an Euclidean time discretization, also $N_\tau$ has to be taken under control in estimating observables correctly. In Fig.~\hyperref[fig:stiff]{\ref{fig:stiff}(b)}, it can be seen that the underestimation of $N_\tau$ implies an underestimation of $\Upsilon$ that converges to its actual value only for $N_\tau\rightarrow\infty$. The underestimation is severe especially at low temperatures and in the critical region. Actually, in the case of Fig.~\ref{fig:stiff} convergence near the critical point is only achieved for $N_\tau=1000$, whereas at lower temperatures it even remains insufficient.

The peculiar behaviour of the imaginary time convergence, $N_\tau$, could have been the origin of the re-entrant phase transition claimed for similar models \cite{Capriotti2003}. Our results clearly demonstrate that such a re-entrant phase transition does not exist in our model at low temperatures. Indeed this observation points out that reliable results, based on Monte Carlo methods, are extremely demanding from the computational point of view especially at low temperatures. Indeed, a different computational approach will be introduced in the next section (\ref{sec:finite-size-scaling}) in order to obtain reliable zero temperature values of the stiffness.

\subsection{\label{sec:finite-size-scaling}Finite size scaling}
The aim of this section is to show how to get rid of finite size effects in estimating the critical temperature from the stiffness data obtained via Monte Carlo. In the subsubsection {\ref{sec:BKT} we discuss the finite temperature BKT transitions while in the subsubsection {\ref{sec:3d-xy} we analize the zero temperature quantum phase transition.  Finally, in subsubsection \ref{sec:fake_3d}, we discuss the role of quantum fluctuations at finite temperature close to the quantum critical point.

\subsubsection{\label{sec:BKT}Finite temperature phase transition (BKT)}

Following the approach proposed in Refs.[\onlinecite{boninsegni,Hsieh2013}], first, we calculate $T_c(L)$ solving the equation 
\begin{equation}
    \Upsilon_L(U, T_c(L)) = \frac{2T_c(L)}{t\pi},
    \label{eq:stiff-scal}
\end{equation}
that is based on the finite $L$ version of Nelson-Kosterlitz (NK) criterion \cite{nk-criterion} providing the universal critical jump in the limit $L\mapsto\infty$. Then
we use the following finite size scaling
\cite{boninsegni, Hsieh2013} based on a renormalization group approach:
\begin{equation}
    T_c(L)=T_c(\infty)+\frac{a_T}{\ln^2(b_T L)}\label{eq:T-BKT}.
\end{equation}

In Fig.~\ref{fig:finite size scaling} we show the fitting for $U/t=0.1$ and $U/t=4$. The systematic application of the method allowed us to build up the phase diagram of Fig~\ref{fig:diagram}. It is worth noting that some of the points shown in the Fig~\ref{fig:diagram} (black triangles) have been obtained keeping the temperature fixed and scanning the stiffness as a function of $U$. In the latter case the scaling becomes:

\begin{equation}
    U_c(L)=U_c(\infty)+\frac{a_T}{\ln^2(b_T L)}\label{eq:U-BKT}.
\end{equation}
having extracted $U_c(L)$ from Eq.~\eqref{eq:stiff-scal} (see Fig.~\hyperref[fig:finite size scaling]{\ref{fig:finite size scaling}c} and~\hyperref[fig:finite size scaling]{\ref{fig:finite size scaling}d}).

\begin{figure}[!ht]
    \centering
    \includegraphics[width=\linewidth]{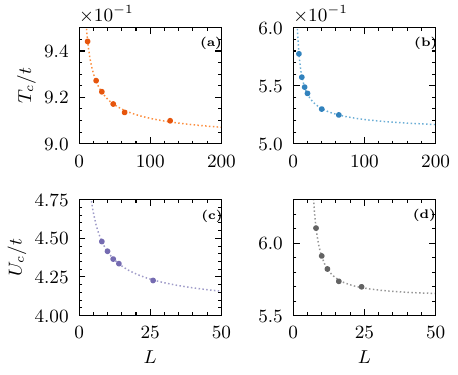}
    \caption{Finite size-scaling for different values of $U$, $T$, and  $N_\tau$: (a) $U/t=0.1$, $N_\tau=100$; (b) $U/t=4$, $N_\tau=1000$; (c) $T/t=0.5$, $N_\tau=1000$; (d) $T/t=0.05$, $N_\tau=5000$. The BKT scaling is valid for both directions, $T$ and $U$, of the phase diagram reported in Fig~\ref{fig:diagram}.}
    \label{fig:finite size scaling}
\end{figure}

\subsubsection{\label{sec:3d-xy}Zero temperature phase transition}

Despite the critical behaviour at finite temperature, in the thermodynamic limit $L\rightarrow\infty$ and  at $T=0$, the model undergoes a new phase transition which is not any longer BKT-type but acquires a strictly quantum nature. It is well known that it falls in the 3D-XY~\cite{Girvin-3D, 3d-xy} universality class where the helicity modulus size scaling law~\cite{finite-size-scaling}, near the critical point, is
\begin{equation}
    \Upsilon=L^{-z}f\left((U-U_\text{c})L^{\frac{1}{\nu}}\right).
    \label{eq:3d-scaling}
\end{equation}
The dynamic exponent is $z=1$~\cite{3Dcritical} and the correlation length exponent is $\nu=0.671$~\cite{nu3d}, such that at the critical point $\Upsilon(U_{\text{c}},L)\cdot L=\text{cost}$.

Exploiting the fact that $z=1$, we can reach simultaneously $L\mapsto\infty$ and $\beta\mapsto\infty$ ($T=0$) keeping $L=\beta$. In order to reduce the computational effort in achieving reliable values for the stiffness at low temperatures, we first perform a Monte Carlo run at finite $\Delta\tau$ for different values of $L$ and $N_\tau=L/\Delta\tau$, then we exploit the scaling law \eqref{eq:3d-scaling} and get an estimation of the critical value $U_c$ at that finite $\Delta\tau$. Finally, we repeat the procedure by progressively reducing $\Delta\tau$ to estimate the asymptotic value.

\begin{figure}[!ht]
    \centering
    \includegraphics[width=\linewidth]{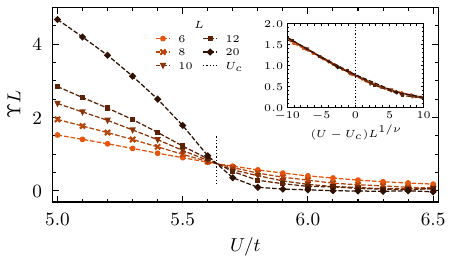}
    \caption{3D-XY scaling of the stiffness $\Upsilon$ vs interaction energy $U$ with $N_\tau=150\beta$, with $\beta=L$ and $U_\text{c}/t=5.63(2)$. In the inset, the data collapse with $\nu=0.671$ is showed.}
    \label{fig:3Dstiff}
\end{figure}

Fig.~\ref{fig:3Dstiff} shows the 3D-XY scaling of $\Upsilon L$ for $\Delta\tau = 1/150$ and different values of $L$. The curves at different sizes intersect all at the same fixed point that, according to Eq.~\eqref{eq:3d-scaling}, signals the critical point $U_c$. Using this value of $U_c$ in Eq.~\eqref{eq:3d-scaling}, we can see that that the data collapse takes place (inset of Fig.~\ref{fig:3Dstiff}). Since the actual critical value is reached as $N_{\tau}\to \infty$, we repeated the procedure for progressively larger values of $N_{\tau}$ (see Fig.~\ref{fig:iso3D}). From the evident linear relationship between the estimated $U_c$ values as a function of $\Delta\tau$, it was possible to extract the limit value from a linear fit $U_c(\Delta\tau)/t=U_c(0)/t+a\Delta\tau$, obtaining $U_c(0)=5.72(1)$.
\begin{figure}[!ht]
    \centering
    \includegraphics[width=\linewidth]{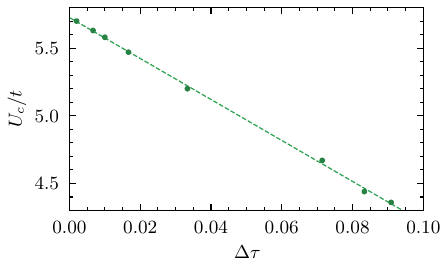}
    \caption{Variation of the 3D-XY fixed point as a function of $\Delta\tau$. In the limit $\Delta\tau\rightarrow0$, the system converges to the critical value $U_c(0)/t=5.72(1)$. This value is obtained by the data fit $U_c(\Delta\tau)/t=U_c(0)/t+a\Delta\tau$.}
    \label{fig:iso3D}
\end{figure}

It should be noted that the estimated value for $U_c$ is significantly greater than the value of $U_c/t=4.25(2)$ previously reported in the literature~\cite{Jiang2022,Cha,Girvin-3D}. The difference is due to the coupling anisotropy, $K_\tau\ne K_s$,  of our model action (see Eq. (\ref{eq:action}))  that is generally overlooked in other papers. Indeed, if we are only interested in the universal critical properties, such as critical exponents, the anisotropy  can be safely neglected and we can simplify the model taking $K_\tau= K_s=(t/U)^{0.5}$. The latter choice corresponds to selecting $\Delta\tau=(t U)^{-0.5}$. On the other hand, the anisotropy matters for the non universal critical quantities as the critical value $U_c$. In order to get it, we have kept the anisotropy performing the limit $\Delta\tau\mapsto 0$ (see Fig.~\ref{fig:iso3D}).

\subsubsection{\label{sec:fake_3d}Quantum fluctuations at finite temperature}
We have shown that the critical line that separates the superconducting phase from the other phases is of BKT nature  except at $T=0$ where a truly quantum transition occurs (see green squares and the orange circle in Fig.~\ref{fig:diagram}). The quantum critical point is in the same universality class of the classic 3D-XY model. This singular behaviour of the $T=0$ quantum phase transition suggests that, even at finite temperatures, significant quantum fluctuations could be observed. In order to clarify this point, we have computed $\Upsilon \cdot L$ as function of $U$ at constant $T$. The results are reported in Fig.~\ref{fig:fake3D} for different values of $L$. For small $L$, the data show a common point as expected for the $T=0$ transition, however, this is not any longer true at larger values of $L\ge20$. We interpret this result as an effect of strong quantum fluctuations that would trigger a 3D-XY critical transition if the correlation length in the imaginary time could reach the limiting value $\beta\mapsto\infty$. Actually, the finite value of $\beta$ frustrates this possibility.
 
%For small sizes ($L=8,10,12$), the stiffness multiplied by $L$ exhibits a well-%defined crossing point, consistent with the 3D-XY scaling behavior. For larger sizes, %however, the phase stiffness no longer crosses at the same fixed point, indicating a %deviation from the expected scaling and signaling that the 3D-XY transition is no %longer possible.

\begin{figure}[!ht]
    \centering
    \includegraphics[width=\linewidth]{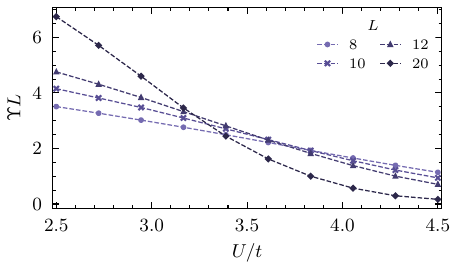}
    \caption{Stiffness $\Upsilon$ as a function of $U/t$  for $N_\tau=10^3$, $T/t=0.8$, and several system sizes $L$.}
    \label{fig:fake3D}
\end{figure}

\subsection{\label{sec:vortex}Vortex density}
A crucial role in BKT transitions is played by vortices that are the main excitations of the system. In order to better understand quantum effects, we calculate the density of vortices as a function of $T$ and $U$. The definition of the local vorticity \cite{Kosterlitz1974} is
\begin{equation}
    \oint_\gamma d\phi(r) = 2\pi\sum_{l\in\gamma} q_l,
\end{equation}
where the path $\gamma$ is tipically a plaquette,  with $q_l\in\mathbb{Z}-\{0\}$ the charge of the vortex. Assuming that the main vortices proliferating near the BKT phase transition carry unitary charge ($|q|=1$), the total vortex density on a discrete lattice is obtained by summing the phase differences over all plaquettes, hence
\begin{equation}
    n_v=\frac{1}{2\pi N_\tau L^2}\sum_{\tau,i} 
    \left|
    \sum_{k\in\square_i}\text{mod}_{2\pi}{\left[\Delta\phi_k(\tau)\right]}
    \right|,
\end{equation}
where $k$ is the index of the bond of the single plaquette with root site $i$ (denoted by $\square_i$), $\Delta\phi_k$ is the phase difference between the site phase connected by the bond, and $\text{mod}_{2\pi}: [-2\pi, 2\pi]\rightarrow[-\pi,\pi)$ \cite{vortex} is defined as
\begin{equation}
    \text{mod}_{2\pi}{\left[\Delta\phi\right]} = \Delta\phi-2\pi\lfloor 1/2-\Delta\phi/2\pi\rfloor
\end{equation}

\begin{figure}[!ht]
    \centering
    \includegraphics[width=.45\textwidth]{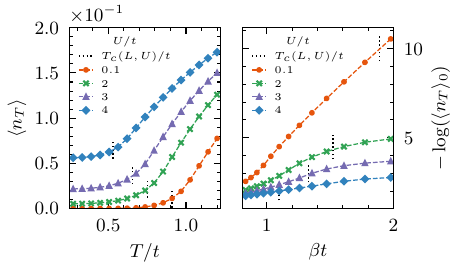}
    \caption{The total number of vortices $n_T$ as a function of the temperature $T$ (left panel) and  $\beta$, the inverse of the temperature, (right panel) for several values of $U/t$.}
    \label{fig:nv_figure}
\end{figure}
In the classical limit — consistent with $U/t=0.1$ — the number of vortices $n_T$ below the critical temperature is small and goes to zero exponentially with the temperature. In the quantum regime, however, the coupling $U$ is a new mechanism capable of  exciting vortices in addition to the thermal mechanism. In fact, we observe that the number of vortices increases with $U$ and does not approach a zero value at low temperatures (see Fig.~\ref{fig:nv_figure}). However, above the critical temperature, $n_T$  exhibits the same exponential growth as in the classical regime. We interpret these results in the following way: quantum fluctuations, although able to change the number of vortices in the system, do not change the nature of the transitions that remain BKT as already observed in Fig 1. The only effect is observed in the value of $T_c$ that decrease with $U$. Quantum fluctuations are able to screen the vortex-antivortex interaction but not to modify their logarithmic form that is the finger-print of the BKT transition.
Note that we calculate the density of the total number of vortices, $n_T$, without distinguishing between bound vortex-antivortex pairs, $n_b$, and unbound free vortices, $n_f$. Only in the superconducting phase, below $T_{BKT}$, $n_T=n_b$.

\subsection{\label{sec:corr-len}Correlation length and Resistance}

The temperature dependence of the correlation length, $\xi$, allows us to further characterize the system. Beside signaling the critical temperature, $\xi$ is involved in the resistance response of the system. Indeed, at $T\ne 0$, the onset of a resistive phase is due to the formation of free vortices that dissipate energy when interacting with an external current~\cite{Minnhagen_rev}. Then, above BKT transition, that is characterized by unbinding of vortex-antivortex pairs, there is the generation of a free vortex density, $n_f$, that, in turn, is proportional to $1/\xi^2$~\cite{Benfatto2009}.  As shown by Halperin \textit{et al.}~\cite{Halperin1979}, the temperature dependence of the correlation length near the BKT transition and, then, the resistance are given by the following Halperin and Nelson law %~\eqref{eq:HN_resistance}
\begin{equation}
    \frac{R_{s}}{R_N}\propto 1/\xi^2_+ \propto 1/\xi^2_c\exp{\left[-B/\sqrt{T-T_c}\right]},
    \label{eq:HN_resistance}
\end{equation}
where $R_N$ is the resistance at $T\gg T_c$, that is in the normal (hence the subscript $N$) phase. 

We assume that the resistance in our model is always related to the presence of unbounded  vortices  except at $T=0$. As we have shown in section \ref{sec:3d-xy}, at $T=0$ a different, truly quantum, phase transition sets in associated to an effective 3D nature. In this situation, simple vortex-antivortex pairs are not any longer the main excitations and the resistance should not be proportional to $n_f$ and, therefore, to $\xi^{-2}$.

We are able to obtain the temperature dependence of the resistance for different values of $U$ in a numerically exact way by using Monte Carlo approach. Under the assumption that the correlation function decays as $\exp(-r/\xi)$, a good approximation for the correlation length is given by~\cite{corr.length}
\begin{equation}
    \xi = \frac{1}{2\sin(\pi/L)} \sqrt{\frac{\langle m(\vec{0})^2 \rangle}{\langle m(\vec{k}_0)^2 \rangle} - 1},
\end{equation}
where $m(\Vec{k})$ is the discrete Fourier transform of the magnetization calculated at zero momentum and at $\Vec{k}_0=(2\pi/L, 0)$. 

Fig.~\ref{fig:xiHN} shows the Monte Carlo results for different values of $U$. The Halperin and Nelson law~\eqref{eq:HN_resistance} remains valid for all the studied values of $U$ smaller than the critical interaction energy  and for a significant range of temperatures above the critical temperature. Somehow the only effect of $U$ is to change the critical temperature. In fact, if the Halperin and Nelson law ~\eqref{eq:HN_resistance}  is used to fit the Monte Carlo data, the fit parameters $\xi^2_c$ and $B$ are independent of $U$.
Therefore, when the data are plotted as a function of $T-T_c(U)$, we observe a  data collapse (see inset of Fig.~\ref{fig:xiHN}) involving  that, for all the values $U$,  the same mechanism controls the resistance.  As a result, the logarithm of $\xi^{-2}$ diverges  as $1/\sqrt{T-Tc}$ in the thermodynamic limit. In Fig.~\ref{fig:xiHN}, near the critical temperature, the data deviate from the asymptotic behavior due to size effects~\cite{Benfatto2009}. Indeed, when the correlation length, that diverges at the critical temperature in the thermodynamic limit, becomes close to the finite size $L$ used in the Monte Carlo simulations, we clearly observe a tendency to saturation.

\begin{figure}[!ht]
    \centering
    \includegraphics[width=\linewidth]{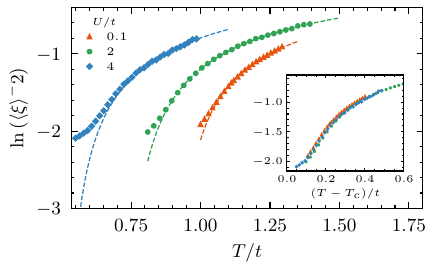}
    \caption{Logarithm of the inverse square correlation length as a function of the temperature $T$ for $L = 48$ and $U/t = 0.1, 2, 4$. The dash lines are obtained using the Halperin and Nelson law~\eqref{eq:HN_resistance}  to fit the Monte Carlo data, using as fit parameters $1/\xi^2_c$ and $B$. 
    In the bottom-right inset, we show the data collapse obtained  when the data are plotted as a function of $T-T_c(U)$.  }
    \label{fig:xiHN}
\end{figure}
The scenario changes drastically at $U>U_c$. In this region the system is any longer critical at any temperature and we always observe  a finite resistance. In particular, the resistance exhibits a minimum, as shown in Fig.~\ref{fig:xi-i}, at values of $U$ that move towards $U_c$ approaching $U_c$. This type of behaviour in the resistance can be associated  with a finite temperature MI crossover. In Fig.~\ref{fig:diagram} the crossover between metallic and insulating phases are identified by blue points. 
\begin{figure}
    \centering
    \includegraphics[width=\linewidth]{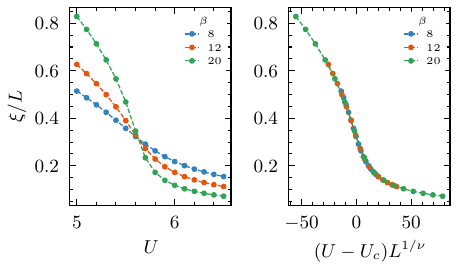}
    \caption{Correlation length calculated for $L=\beta$ and $N_\tau=150L$. According to the finite-size scaling hypothesis $\xi = L\,g\left((U-U_c)L^{1/\nu}\right)$, the ratio $\xi/L$ exhibits a fixed point at $U_c$. The observed data collapse when plotted against the argument of $g$ confirms that the correlation length follows 3D-XY scaling.}
    \label{fig:scaling_xi}
\end{figure}

This analysis would suggest that the MI crossover line would end at $U=U_c$ and $T=0$, where critical behaviour of $\xi$ should become a power law as predicted in the 3D-XY model (see Figure~\ref{fig:scaling_xi}). At this point it has been claimed that the resistance  should reach the quantum limit \cite{Sachdev} becoming universal. Even if there is no general consensus on this claim~\cite{Girvin-3D}, our results suggest that $(1/\xi)^{-2}\mapsto 0$ and, then, also $R\propto  n_f\propto  (1/\xi)^{-2}$ should reach a vanishing value. However, at the critical point, as discussed at the beginning of this section, we expect that no longer $R\propto n_f$ since the mechanism controlling the phase transition is not a simple unbinding of vortex-antivortex pairs as in the $BKT$ regime.

\begin{figure}[!ht] 
    \centering
    \includegraphics[width=\linewidth]{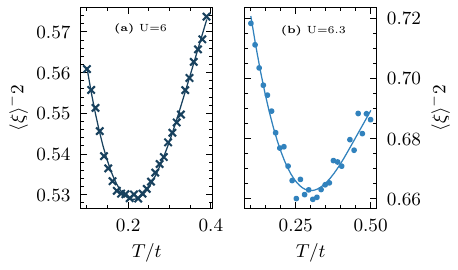}
    \caption{Inverse square correlation length as a function of the temperature. (a) Plot for $U/t = 6$, $L = 32$, and $N_\tau = 7000$; (b) Plot for $U/t = 6.3$, $L = 32$, and  $N_\tau = 7000$. This behavior can be associated with the M-I finite temperature crossover.
}
    \label{fig:xi-i}
\end{figure}

\section{Conductivity properties}
In this section, we extend the Monte Carlo calculation to the transport properties resorting to linear response theory (for details see Appendix B). First, in subsection \ref{sec:drude}, we discuss the Drude weight, then, in subsection \ref{sec:regularcon}, the regular part of the conductivity.

In order to compute the real part of the conductivity response function defined as~\cite{shastry, evertz}
\begin{equation}\label{eq:re-cond}
    \Re\left\{ \sigma(\omega, T) \right\} = D(T)\delta(\omega)+\sigma_{\text{reg}}(\omega, T),
\end{equation}
where $D(T)$ is the Drude weight, we compute via Monte Carlo method  the imaginary time current-current correlation function
\begin{equation}
   F(\tau)= \langle I(\tau)I(0)\rangle,
\end{equation}
where $I(\tau) = e^{\tau H } I e^{-\tau H}$ is the imaginary-time Heisenberg evolution of $I$ given in Eq.~\eqref{eqn:current_operator} which, in the path integral formulation, takes the form  \eqref{eq:path-cur}.
%$I(\tau) = t \sum_{\langle ij\rangle}\sin[\phi_i(\tau)-\phi_j(\tau)]$. 

From the knowledge of $F(\tau)$, we can reconstruct the conductivity on the real frequency axis resorting to analytical continuation techniques \cite{evertz, shastry}). From a computational point of view, evaluating the correlation function $\langle I(\tau)I(0)\rangle$ is considerably more demanding than computing other quantities, as it is not averaged over the imaginary-time degrees of freedom. Achieving results with smoothness and stability comparable to those obtained when averaging over all degrees of freedom requires a computational time at least $N_\tau$ times longer.

\subsection{The Drude weight}\label{sec:drude}
In superconducting systems, the Drude weight consists of two contributions: one, $D_s$, from superconducting charge carriers, and the other, $D_{ns}$ from non-superconducting ones, such that $D=D_s+D_{ns}$. The total Drude weight $D(T)$ is given by (see Appendix \ref{apx:lrt})
\begin{equation}\label{eq:drude}
    D(T)=\left\langle\sum_i\frac{dI}{d\phi_i}\right\rangle+\Pi(i\omega_n\rightarrow0),
\end{equation}
where $\Pi(i\omega_n)$ is the current-current correlation function in Matsubara frequencies $\omega_n=2\pi n/\beta\hbar$. The superconductive Drude weight \cite{scalapino}
\begin{equation}\label{eq:sc-drude}
    D_s(T)=\left\langle\sum_i\frac{dI}{d\phi_i}\right\rangle+\Pi(i\omega_n=0)
\end{equation}
in our units ($\hbar = 2e = k_B=1$) coincides with the definition of stiffness~\eqref{eq:stiffness} \cite{scalapino}. Therefore, the difference between $D$ and $D_s$ lies in the difference between the $\Pi(i\omega_n=0)$ and $\Pi(i\omega_n\rightarrow0)$. It can be shown  \cite{shastry} that
\begin{equation}
    \Pi(i\omega_n\rightarrow0)-\Pi(i\omega_n=0) = \frac{\beta}{Z}\sum_{\substack{n,m\\E_n=E_m}}e^{-\beta E_n}|\braketmatrix{n}{I}{m}|^2.
\end{equation}
This difference is, therefore, linked to the existence of non zero average values of the current on degenerate states.
In the present model, there are no charge carriers other than Cooper pairs, so $D$ should be equal to $D_s$. To prove it, we exploit the Lorentzian sum series form of the $\Pi(i\omega_l)$  \cite{evertz}
\begin{equation}
    \Pi(i\omega_l)=\sum_j c_j\frac{\Delta_j}{\omega_l^2+\Delta_j^2},
\end{equation}
to fit the Fourier transformation of $\left\langle I(\tau)I(0)\right\rangle$. As shown in Fig.~\ref{fig:piwl}, the first two terms of Eq.~\eqref{eq:sc-drude} already converge at the value $\Pi(0)$ 
%=D_s(T)-\left\langle dI/d\phi\right\rangle$ 
for each temperature. This is in agreement with $\Pi(i\omega_n=0)=\Pi(i\omega_n\rightarrow0)$.
\begin{figure}
    \centering
    \includegraphics[width=\linewidth]{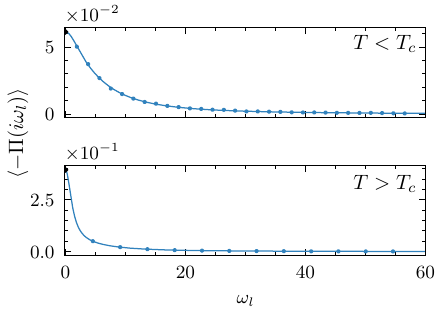}
    \caption{Fourier transform of the imaginary time current-current correlation function (in units of $t$) 
    %{\color{blue}
    %[F.G.C.: o $t(2e/\hbar)^2$, da scegliere cosa scrivere]} 
    as a function of Matsubara frequencies (in units of $t$) for $U/t=4$ and $L=48$. In the upper panel, $T<T_c$ ($T=0.3$), in the lower panel, $T>T_c$ ($T=0.72(4)$). The zero frequency value is indicated by a black diamond point in both panels.  
    %is $\Pi(0)=D_s(T)-\left\langle dI/d\phi\right\rangle$
    }
    \label{fig:piwl}
\end{figure}

\subsection{The regular part}\label{sec:regularcon}
As mentioned above, since our Monte Carlo simulations do not involve real times but only imaginary times, to obtain the conductivity at real frequencies, we use the relationships that link the current-current imaginary time correlations to the regular part of the conductivity (see Appendix \ref{apx:lrt}): 
\begin{equation}
    \langle I(\tau)I(0) \rangle =
    \frac{1}{\pi}\int_0^{+\infty}d\omega\  \sigma_\text{reg}(\omega) D_\omega(\tau),
    \label{eq:ItI0_int}
\end{equation}
where $D_\omega(\tau)=\omega\cosh{[\omega(\beta-2\tau)/2]}/\sinh{(\omega\beta/2)}$ is the bosonic propagator. To find the solution of the integral Eq.~\eqref{eq:ItI0_int} is not simple, since the term ($\langle I(\tau)I(0) \rangle$  is known through data that are affected by statistical fluctuations. In order to avoid non physical solutions, we made a physical ansatz for $\sigma_\text{reg}(\omega)$. We assumed that, for $T>T_c$, in the metallic phase, it is given by the sum of a Lorentzian curve (the characteristic Drude term) and a symmetrized Gaussian curve, therefore
\begin{equation}
    \sigma^+_\text{reg}(\omega)=\frac{A_+}{B_+^2+\omega^2}+C_+\left(e^{-\tau_+^2(\omega-\omega_+^*)^2}+e^{-\tau_+^2(\omega+\omega_+^*)^2}\right),
\end{equation}
while, for $T<T_c$, we consider only a symmetrized  Gaussian curve, hence
\begin{equation}
    \sigma^-_\text{reg}(\omega)=C_-\left(e^{-\tau_-^2(\omega-\omega_-^*)^2}+e^{-\tau_-^2(\omega+\omega_-^*)^2}\right),
\end{equation}
where $A_\pm$, $B_\pm$, $C_\pm$, $\tau_\pm$ and $\omega_\pm^*$ are chosen to minimize

\begin{equation}
   \int_0^{\beta/2} \left[\langle I(\tau)I(0) \rangle -
    \frac{1}{\pi}\int_0^{+\infty}d\omega\  \sigma_\text{reg}(\omega) D_\omega(\tau)\right]^2 d\tau
    \label{eq:ItI0_int_square}.
\end{equation}
In order to reduce the free parameters, we minimize Eq.~\eqref{eq:ItI0_int_square} under the constrain  that the temperature dependence of $\sigma^+_{reg}(0)$ is consistent with the Monte    Carlo results for $\xi^{2}$ i.e.,  $\sigma^+_{reg}(0)=K \xi^{2}$. In such a way the minimization provides the value of the constant $K$.

In Fig. 11, we report the zero-frequency regular conductivity (orange crosses) as a function of the temperature at $L=48$. We also report the superfluid stiffness (blue diamonds) in order to individuate the critical region where we expect severe size effects (see Figure \ref{fig:xiHN} and discussion at the end of the section). For temperatures $T < T_c$, the regular part is extremely small implying that the only significant contribution to the zero frequency conductivity stems from the Drude term above $T_c$. The continous orange line is the Halperin--Nelson law given in Eq.~\eqref{eq:HN_resistance}.

In Fig.~\ref{fig:sreg-high}, the results of the minimization for $\sigma^+_\text{reg}(\omega)$ at $T>T_c$ (normal metallic phase) are shown for (a) $U/t=4$ and (b) $U/t=1$. As expected, the metallic behaviour is evidenced by a Drude peak at $\omega=0$ strongly dependent on $U$. However, in spite of what occurs in the classic case ($U=0$), an extra peak appears at finite frequency, i.e., at $\omega/t\simeq 5$ for $U/t=4$ and at $\omega/t\simeq 1$ for $U/t=1$. We ascribe these peaks to the effects of quantum fluctuations since they disappear at $U/t=0$ and are centered at frequencies $\omega\approx U$. In order to make clearer the extra peak contribution, in the inset we plot only the gaussion contribution. We also note that, for $U/t=1$, it is not possible to resolve the two peaks because the distance between the two centers is less than the amplitude of the Gaussian curves, so that the resulting function is zero-peaked.

Below the critical temperature, in the superconductive phase, the regular conductivity $\sigma^-(\omega)$ has a different behaviour. In the classical limit ($U/t=0$) a zero value of the regular conductivity is expected \cite{Halperin1979}. Instead quantum fluctuations are responsible for the peaks at frequencies similar to those obtained in the metallic phase. As shown in Fig.~\ref{fig:sreg-low},  a single peak is observed at different values of $U/t$. More precisely, the regular conductivity $\sigma^-(\omega)$ is characterized by only the symmetrized-Gaussian term centered in $\omega_0/t=5$ at $U/t=4$ (left panel), in $\omega_0/t=1$ at  $U/t=1$ (right panel).
A more systematic analysis is in progress and will be published further. 

We conclude this section discussing briefly the limitations of the analytical continuation techniques. We got a physically sensible result only far enough from the critical temperature (see light-green area in Fig.~\ref{fig:sigma0} for $U/t=4$ and $L=48$). This is due to finite size effects that, in a temperature range around $T_c$, are significant and prevent the analytical continuation to provide physically sounded results. 
%Actually, in the intermediate temperature range, we obtain a Lorentzian term that %decreases moving towards $T_c$ instead to increase according to Halperin-Nelson %formula.

\begin{figure}
        \centering
        \includegraphics[width=\linewidth]{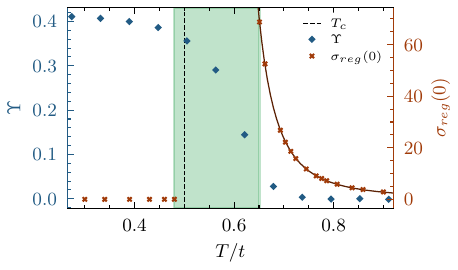}
        \caption{Stiffness and zero--frequency regular conductivity as a function of the temperature $T$ for $L = 48$, $N_\tau = 1000$, and $U/t = 4$. The left vertical axis shows the scale for the stiffness $\Upsilon$, represented by blue diamonds, the right vertical axis that for the zero-frequency conductivity $\sigma_{reg}(0)$, represented by orange crosses and extracted through the inversion of Eq.~\eqref{eq:ItI0_int}. The shaded green region denotes the temperature range affected by finite size effects. For temperatures $T > T_c$ and outside the finite-size-dominated region, $\sigma_{reg}(0)$ decays exponentially following the Halperin--Nelson law given in Eq.~\eqref{eq:HN_resistance} (continuous orange line).}
        \label{fig:sigma0}
    \end{figure}

\begin{figure}
    \centering
    \includegraphics[width=\linewidth]{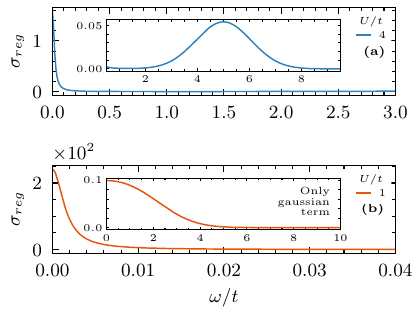}
    \caption{Regular part $\sigma^+_\text{reg}$ 
    %of the 
    %{\color{blue}[F.G.C.: l'unità è $(2e/\hbar)^2=1$, da scrivere?)]}  
    as a function of the angular frequency $\omega$ for $T>T_c(U)$. (a) Plots for $U/t=4$, $L=48$ and $N_\tau=1000$. (b) Plots for $U/t=1$, $L=48$ and $N_\tau=500$. In the insets, only the Gaussian terms are plotted.}
    \label{fig:sreg-high}
\end{figure}

\begin{figure}
    \centering
    \includegraphics[width=\linewidth]{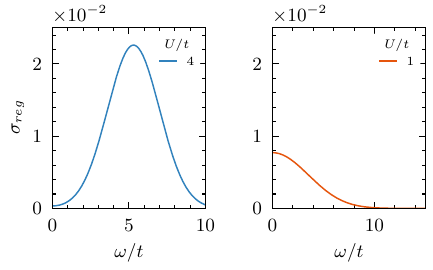}
    \caption{Regular conductivity $\sigma^-_\text{reg}$
    %{\color{blue}[F.G.C.: l'unità è $(2e/\hbar)^2=1$, da scrivere?]} 
    as a function of the angular frequency $\omega$ for $T<T_c(U)$ at $U/t=4$ (left panel) and  $U/t=1$ (right panel). }
    \label{fig:sreg-low}
\end{figure}

\section{\label{sec:conclusions} Conclusions and Discussions}

Our quantum Monte Carlo study of the 2D QXY model has shown that quantum and thermal fluctuations strongly suppress superconductivity. The temperature–interaction phase diagram reveals a clear quantum critical point without reentrant behavior. The critical thermal scaling of the resistance, given by the Halperin and Nelson law, persists up to the critical interaction, while stronger interactions lead to an insulating phase. Finally, the extracted frequency-dependent conductivity displays a finite-frequency response consistent with quantum phase fluctuations, offering a unified picture of superconductivity’s breakdown in two dimensions.

The observation of a superconducting phase at the interface of different oxides, such as LaAlO$_3$/SrTiO$_3$ or LaAlO$_3$/KTaO$_3$, has attracted the attention of many research groups. In particular, the resistance data show a transition from a MS  to MI when the gate voltage across the interface is reduced~\cite{Mallik2022}.  Indeed the gate voltage controls both the charge density and the value of interaction energy $U$ in a way that, unfortunately, is not simple to disentangle from experimental data. On the other hand, one of the main results of our paper is to show that the QXY model indeed exhibits both MS transition and MI crossover depending on the value of $U/t$. The QXY model, then,  provides, in our opinion, a significant contribution to ascertain if quantum fluctuations are relevant to explain the properties of these oxide interfaces. This suggestion should be contrasted with a different point of view that attributes to strong inhomogeneities the primary role for the crossover from MS to MI transition~\cite{Caprara,Benfatto2009}. Another relevant result of our work is the existence of a wide peak in the optical absorption at frequencies in the range of $U$ due to quantum fluctuations. This peak is predicted to be present both below and above the critical temperature. We believe that experimental works, specifically focused on the optical conductivity at finite frequency,  could be very enlightening in order to study the role of quantum fluctuations in these interfaces.

Possible future directions include investigating QXY systems in the presence of Dzyaloshinskii–Moriya interactions, which may induce topological phase transitions~\cite{Rubinstein,Fara}  and originate from spin–orbit coupling in the underlying fermionic or bosonic Hamiltonian~\cite{Perroni1}. It would also be of interest to move beyond the linear-response regime and address the full non-equilibrium dynamics of the system, for example within the Lindblad framework~\cite{Perroni3,Perroni4}.

\section*{Acknowledgment}
G.D.F. acknowledges financial support from PNRR
MUR Project No. PE0000023-NQSTI. C.A.P. acknowledges funding from IQARO (Spinorbitronic Quantum Bits in Reconfigurable 2DOxides)
project of the European Union’s Horizon Europe research and innovation programme under grant agreement
n. 101115190. G.D.F. and C.A.P. acknowledge
funding from the PRIN 2022 project 2022FLSPAJ “Taming Noisy Quantum Dynamics” (TANQU). C.A.P. acknowledges funding from the PRIN 2022 PNRR project
P2022SB73K “Superconductivity in KTaO3 Oxide-2DEG
NAnodevices for Topological quantum Applications”
(SONATA) financed by the European Union - Next Generation EU

\appendix
\section{\label{apx:part.func}Partition function}
The operators $\phi_j$ and $n_k$ satisfy the commutation relation
\begin{equation}
    \left[\phi_j, n_l\right] = i\delta_{jl},
\end{equation}
where $j$ and $l$ denote the site indices, and $n_l = -i \,\frac{\partial}{\partial\phi_l}$ is the number operator.
From this relation, the overlap between the corresponding eigenstates is given by
\begin{equation}
    \braket{n_l}{\phi_j} = \frac{1}{\sqrt{2\pi}}\exp\left\{ in_l\phi_j \right\}\delta_{lj}.
    \label{eq:prod-nphi}
\end{equation}
The partition function is defined as
\begin{equation}\label{eqn:partition}
    Z = \Tr \{ e^ {-\beta H} \}  = \int \D[\{\phi\}] \braketmatrix{\{\phi\}}{e^{-\beta H}}{\{\phi\}},
\end{equation}
where $\ket{\{\phi\}} \equiv \bigotimes_i \ket{\phi_i}$ and $\int {\cal D}[\{\phi\}] \equiv \int \prod_i d\phi_i$.
We discretize the imaginary time by defining $\Delta \tau = \beta/N_\tau$, so that  
\begin{equation}
    \exp\left\{-\beta H\right\}=\left[ \exp\left\{-\frac{\beta}{N_\tau} H\right\} \right]^{N_\tau}.
\end{equation}
Inserting $N_\tau-1$ resolutions of the identity in the form
\[
\int {\cal D}[\{\phi(\tau_k)\}]\ket{\{\phi(\tau_k)\}}\bra{\{\phi(\tau_k)\}}, \qquad k = 1,\dots,N_\tau-1,
\]
into Eq.~\eqref{eqn:partition}, we obtain
\begin{equation}
    \begin{gathered}
    Z = \int {\cal D}[\{\phi\}]\braketmatrix{\{\phi\}}{e^{-\beta H}}{\{\phi\}}=\\
    \int\D[\{\phi(\tau_0)\}]\ ...\int\D[\{\phi(\tau_k)\}]\ ...\int\D[\{\phi(\tau_{N-1})\}]\\
    \braketmatrix{\{\phi(\tau_0)\}}{e^{-H\Delta\tau}}{\{\phi(\tau_1)\}}\cdot...\\
    \cdot\braketmatrix{\{\phi(\tau_k)\}}{e^{-H\Delta\tau}}{\{\phi(\tau_{k+1})\}}\cdot...\\
    \cdot\braketmatrix{\{ \phi(\tau_{N-1})\}}{e^{-H\Delta\tau}}{\{\phi(\tau_0)\}} = \\
    \int \prod_{k=0}^{N-1} {\cal D}[\{\phi(\tau_k)\}] \prod_{k=0}^{N-1} \braketmatrix{\{ \phi(\tau_{k})\}}{e^{-H\Delta\tau}}{\{\phi(\tau_{k+1})\}}
    \end{gathered}
\end{equation}
where $\{\phi(\tau_0)\}\equiv \{\phi\}$, $\tau_k = k\Delta\tau$ and $\tau_N = \tau_0$.
For simplicity, we consider now a single overlap. Using the the Suzuki-Trotter approximation~\cite{suzuki}, we get
\begin{equation}
    \begin{gathered}
        \braketmatrix{\{\phi(\tau)\}}{e^{-H\Delta\tau}}{\{\phi(\tau+\Delta\tau)\}}=\\
        \exp\left\{t\Delta\tau \sum_{\langle i,j \rangle} \cos{ \left( \phi_i(\tau) - \phi_j(\tau+\Delta\tau)\right) } \right\}\cdot\\ \braketmatrix{ \{ \phi(\tau) \} }{ \exp\left\{-\frac{U}{2} \Delta\tau \sum_i n_i^2 \right\} }{ \{ \phi(\tau+\Delta\tau) \} } + O(1/N_\tau^2).
    \end{gathered}
\end{equation}
Introducing a decomposition of the identity in the $n$-basis and using~\eqref{eq:prod-nphi}, the second overlap becomes
\begin{equation}
    \begin{gathered}
        \frac{1}{2\pi} \sum_{ \{n(\tau)\} } \exp\left\{-\frac{U}{2}\Delta\tau \sum_i n_i^2(\tau) \right\}\cdot\\
        \exp\left\{ i\sum_i n_i(\tau)\left[ \phi_i(\tau+\Delta\tau) - \phi_i(\tau) \right] \right\}.
    \end{gathered}
\end{equation}
The resulting sum over $n$ can be evaluated as
\begin{equation}
\begin{gathered}
    \sum_{n} e^{-\frac{U\Delta\tau}{2} n^2+in\phi}=\\
    \sum_{m=-\infty}^{+\infty}\int_{-\infty}^{+\infty}dn\ e^{i2m\pi n +in\phi -\frac{U\Delta\tau}{2} n^2} =\\
    \sum_{m=-\infty}^{+\infty} \sqrt{\frac{2\pi}{U\Delta\tau}}\exp\left\{ -\frac{1}{2U\Delta\tau} (\phi - 2\pi m)^2 \right\} \simeq e^{\frac{1}{U\Delta\tau} \cos{\phi}},
\end{gathered}
\end{equation}
where in the last step we used the Villain approximation~\cite{Villain1975}.
Thus, up to corrections that vanish as $N_\tau \to \infty$, the quantum model is mapped onto an effective classical model with partition function
\begin{equation}\label{eqn:partiton_final}
    Z = \int\D[\{\phi(\tau)\}]\ e^{-S[\{\phi(\tau)\}]} +O(1/N_\tau),
\end{equation}
where the effective action $ S[\{\phi(\tau)\}]$ is given in Eq. (\ref{eq:action}).
%\begin{equation}
%    \begin{gathered}
%        S[{\phi}]=\sum_{\substack{\langle i j \rangle\\ \tau}} K_s \cos{\left(\phi_i(\tau)-%\phi_j(\tau)\right)} + \\
%        \sum_{i,\tau} K_\tau \cos{\left(\phi_i(\tau)-\phi_i(\tau+\Delta\tau)\right)}.
%    \end{gathered}
%\end{equation}
The measure $\int \D[\{\phi(\tau)\}]$ in Eq.~\eqref{eqn:partiton_final} has to be understood as $\int \prod_k {\cal D}[\{\phi(\tau_k)\}]$.
\section{\label{apx:lrt}Linear Response Theory}
From the linear response theory \cite{Mahan}, the conductivity is given by 
\begin{equation}\label{eq:cond}
    \sigma(z)=\frac{i}{z}\left[\Pi(z)+\left\langle\sum_i\frac{dI}{d\phi_i}\right\rangle\right],
\end{equation}
with polarization
\begin{equation}
    \Pi(z)=\int_0^\infty\ dt\ e^{izt}\ \Pi(t) \label{eq:piz},
\end{equation}
and $\Pi(t)=-i\theta(t) \left\langle\left[I(t),I(0)\right]\right\rangle$. Using the Lehmann representation, i.e., expressing the trace sum on the eigenbasis $\{\ket{n} \}$ of $H$, it is straightforward to show that
\begin{equation}
    \begin{gathered}
        \lim_{\epsilon\rightarrow0^+}
        \Re\left[\sigma(z=\omega+i\epsilon)\right] =\\
        =\left[2\pi\sum_{\substack{n,m\\E_n\neq E_m}}c_{nm}+\pi\left\langle\sum_i\frac{dI}{d\phi_i}\right\rangle\right]\delta(\omega)\\
        -\pi\sum_{\substack{n,m\\E_n\neq E_m}}c_{nm}\left[\delta(\omega-\Delta_{nm})+\delta(\omega+\Delta_{nm})\right],
    \end{gathered}
\end{equation}
where $E_n$ is the eigenvalue of the corresponding eigenstate $\ket{n}$, $c_{nm}=p_n|\braketmatrix{n}{I}{m}|^2/\Delta_{nm}$ such that $\Delta_{nm}=E_n-E_m$ and $p_n=e^{-\beta E_n}/Z$. As a result, the Drude weight is given by the zero centred delta coefficient while $\sigma_{\text{reg}}(\omega)$ by the non zero delta linear combination.\\
Defining
\begin{equation}
    \Pi(i\omega_n)=-\int_0^{\beta}d\tau\ e^{i\omega_n\tau}\langle I(\tau)I(0)\rangle,
\end{equation}
it is easy to show that $D(T)-\langle \sum_i dI/d\phi_i\rangle=\Pi(i\omega_n\rightarrow0)$ and thus Eq. (\ref{eq:ItI0_int}), 
with $D_\omega(\tau)=\omega\cosh{[\omega(\beta-2\tau)/2]}/\sinh{(\omega\beta/2)}$ the bosonic propagator, 
%\begin{equation}
%    \langle I(\tau)I(0) \rangle =
%    \frac{1}{\pi}\int_0^{+\infty}d\omega\  \sigma_\text{reg}(\omega) D_\omega(\tau)
%\end{equation}
since, as demonstrated in Section \ref{sec:drude}, $\Pi(i\omega_n\rightarrow0)=\Pi(i\omega_n=0)$.

\bibliography{bibliography}

\end{document}